\documentclass[12pt]{article}
\usepackage{amssymb}

\def\del{\partial}
\def\Ds{D\!\!\!\! /}

\def \q{\mathcal{ Q}}

\def\S{\mathcal{S}}
\def\P{\mathcal{P}}
\def\D{\mathcal{D}}
\def\B{\mathcal{B}}
\def\X{\mathcal{X}}

\def\qu{\mathit{Q}}
\def\qubar{\bar\mathit{Q}}
\def\vD{\vec{D}}
\def\vt{\vec{\tau}}
\def\kok{\sqrt{2}}
\def\yarim{{{1}\over{2}}}

\def\a{\alpha}
\def\b{\beta}
\def\g{\gamma}
\def\d{\delta}
\def\e{\epsilon}
\def\se{\varepsilon}

\def\t{\theta}
\def\k{\kappa}
\def\la{\lambda}
\def\t{\theta}
\def\s{\sigma}

\def\f{\phi}

\def\fa{\Phi^{A}}
\def\fas{\Phi_{A}^{*}}

\def\x{\xi}

\def\adot{\dot{\alpha}}
\def\bdot{\dot{\beta}}

\def\cbar{\bar{c}}

\def\labar{\bar{\lambda}}
\def\sbar{\bar{\sigma}}

\def\xbar{\bar{\xi}}
\def\tbar{\bar{\theta}}

\def\ft{\phi^{\dag}}

\def\se{\mathcal{E}}

\def\be{\begin{equation}}
\def\ee{\end{equation}}
\def\bea{\begin{eqnarray}}
\def\eea{\end{eqnarray}}

\def\kok{\sqrt{2}}
\def\yarim{{{1}\over{2}}}
\def\intx{\int d^4 x}

\begin{document}

%

\thispagestyle{empty}
\begin{flushright}hep-th/0307279\\
\end{flushright}
\vskip5em
\begin{center}

{\Large{\bf N=2 Super Yang Mills Action and BRST Cohomology }}

\vskip1cm
K. \"{U}lker
\vskip2em
{\sl Feza G\"{u}rsey Institute,}\\
{\sl \c{C}engelk\"{o}y, 81220, \.{I}stanbul, Turkey}\\
\vskip1.5em
\end{center}
\vskip1.5cm
%
\begin{abstract}

\noindent The extended  BRST cohomology of N=2 super Yang-Mills theory  is discussed in the framework of Algebraic
Renormalization. In particular, N=2 supersymmetric descent equations are derived from the cohomological
analysis of linearized Slavnov-Taylor operator $\B$. It is then shown that both off- and on-shell N=2 super Yang-Mills
actions are related to a lower-dimensional gauge invariant field polynomial $Tr\f^2$ by solving these descent
equations. Moreover, it is found that these off- and on-shell solutions differ only by a $\B-$exact term, which can be
interprated as a consequence of the fact that the cohomology of both cases  are the same.

\end{abstract}

\vskip2em
\noindent {\small{e-mail: kulker@gursey.gov.tr}}
\vfill
\eject
\setcounter{page}{1}
%
%
\section{Introduction and conclusions}

One of the main reason that supersymmetric quantum field theories have been extensively studied these last years
 is  that  they display important finiteness properties due to the cancellations of ultraviolet divergences
\cite{zumino} that are first verified by using superfield formalism in the superspace \cite{fujikawa}. In component
field  formalism, that is needed when calculations on nontrivial backgrounds are considered,  these non-renormalization
theorems are derived due to the fact that the non-renormalized interaction terms and/or the actions themselves can be
written as multiple supervariations of one chirality of lower dimensional field monomials \cite{flume,sorella,ulker1,
kraus}. The algebraic source of these results are related with the cohomological structure of supersymmetric models.

On the other hand, to construct the exact supersymmetric interaction terms and/or the actions  by
applying super-variations to lower dimensional field monomials\footnote{For a similar approach of constructing N=1
globally and locally supersymmetric actions and also for the discussion of anomalies, see \cite{bra2,bra3}.}, one has to use
off-shell formulation of the supersymmetry, i.e. supersymmetric field content of the theory should include auxiliary
fields \cite{ulker1}. Otherwise, supersymmetry algebra realized without auxiliary fields (on-shell supersymmetry) closes
modulo equations of motion terms and as a consequence the on-shell supervariations of these lower dimensional field
monomials give different expressions then the original interaction terms and actions \cite{ulker1}. For instance when
both of the N=1 and N=2 super Yang-Mills (SYM) actions are derived by using pure on-shell supervariations of lower
dimensional field monomials, the resulting expressions differ from the original one (that can be found by using
off-shell variations) by equation of motion terms and there is not a non-trivial way to restore these missing terms
\cite{ulker1}.

The problem of above mentioned on-shell closure of the algebra can be overcome by extending the BRST transformations to
include supersymmetry transformations \cite{w,mag,mpw} in the algebraic renormalization framework \cite{psbook} that is
structurally equivalent \cite{weinberg} to Batalin-Vilkovisky formalism \cite{batalin}. (Note also that the extension of
BRST transformations that includes arbitrary rigid symmetries is given in Ref.\cite{bra1}.) It is then possible to
derive an off-shell nilpotent Slavnov-Taylor operator $\B$ from an action functional that is extended by antifields
(sources for extended BRST transformation of the fields) both for off- and on-shell supersymmetric cases. However,  for
on-shell supersymmetric case further extension of the action by adding some (none standard) quadratic terms in
the anti-fields of fermions to the action is needed in order to get  such a nilpotent operator \cite{w,mag,mpw}. The
renormalization  program then reduces to an algebraic discussion of the cohomology of a linearized Slavnov-Taylor
operator defined on the space of integrated field polynomials. The counter terms and the possible anomalies are then the
solutions of this cohomology problem with ghost number 0 and 1 respectively \cite{psbook}.

Therefore, since the supersymmetric actions also belong to the cohomology of Slavnov-Taylor operator $\B$ with ghost
number 0, in order to relate the above mentioned actions exactly to lower dimensional field monomials, it is natural to
study a (extended) BRST cohomology problem in the algebraic renormalization framework by using the set of descent
equations. In a recent paper \cite{ulker2} we were able to achieve this goal for both off- and on-shell N=1 SYM
action and our aim in this paper is to extend the analysis of Ref.\cite{ulker2} to N=2 SYM theory.

Our motivation is twofold: first of all,  it is well known that the perturbative beta function of the theory receives
only one-loop contribution. This non-renormalization theorem has been proven algebraically in Ref.\cite{sorella} by
using the above mentioned set of descent equations for the  twisted version of N=2 SYM theory \cite{witten}. The proof
relies on a relationship between the lower dimensional gauge invariant field monomial $Tr\f ^2$ and the twisted action
of N=2 SYM theory (see also Ref.\cite{sorlec}). Therefore, we find also intresting to study the original, untwisted N=2
theory \footnote{Note also that, in a recent paper \cite{ghq} background field method (BFM) has been formulated for the
twisted N=2 SYM. Our results may also be useful to formulate BFM for the orginal, untwisted case.}. As it will be shown
explicitly, solutions of the set of descent equations, that can be derived from the operator $\B$ by using Wess-Zumino
consistency condition, give also a similar relation between both off- and on-shell supersymmetric actions and the gauge
invariant field monomial $Tr\f ^2$, where the scalar field $\f$ is the lowest component of the N=2 vector multiplet. In
other words, we show explicitely that both off- and on-shell actions of the theory can be constructed from $Tr\f^2 $.
 It is then straightforward to generalize the algebraic criterion for the non-renormalization theorem given for the
twisted N=2 SYM \cite{sorella} to both off- and on-shell supersymmetric (untwisted) N=2 SYM theory. Note that, this
result is not surprising since the twisted Yang-Mills theory can be obtained from the original N=2 SYM by directly
performing field redefinitions  (i.e. without twisting) \cite{ulker3}.

Our second motivation is to understand further the structure of the descent equations for globally supersymmetric gauge
theories. The structure of these descent equations are quite different from the ones of nonsupersymmetric
theories and the solutions of these equations are highly constrained due to supersymmetry. In a previous work
\cite{ulker2}, these descent equations are found for N=1 SYM by using the supersymmetric structure of the
theory\footnote{Recently, a systematic framework is proposed in order to solve the supersymmetric descent equations
\cite{sorella2}. The approach given in Ref.\cite{ulker2} for N=1 SYM is also shown to be  consistent with this framework
\cite{sorella2}.}. Therefore, it is also useful to extend the method given in Ref.\cite{ulker2} in order to derive the
complete set of descent equations for (untwisted) N=2 SYM theorem. The structure of these descent equations for N=2 SYM
are similar with that of twisted SYM \cite{sorella,sorlec,sorella2}, as expected . However, it is worth mentioning that 
when these equations are compared with that of N=1 SYM case, it is seen that the structure and also the number of the
descent equations are determined due to supersymmetry together with the corresponding  R-symmetry of the theory. For
instance, as it will be derived explicitly in this paper, the descent consists of five equations (i.e. four descendants)
due to the $SU(2)_R$ symmetry of N=2  SYM theory whereas there was only three (i.e. two descendants) for N=1 case
\cite{ulker2}. This fact may also have intresting outcomes when descent equations for N=4 SYM are considered since the
internal R-symmetry group of the theory is $SU(4)$.

The organization and the results of the paper are as follows. In Sec.II, we review briefly the extension of BRST
transformations to include N=2 supersymmetry and we introduce corresponding Slavnov-Taylor (ST) operator $\B$ for
both off- and on-shell N=2 supersymmetric Yang-Mills theory. In Sec.III, the descent equations, which arise
from the cohomological analysis of $\B$, are derived for N=2 SYM theory. In Sec.IV, by solving these equations
for ghost number 0, we derive algebraic identities that relate the 2-dimensional gauge invariant field monomial $Tr\f^2$
to both off and on-shell supersymmetric (extended) Yang-Mills actions. We also show that these two solutions of off- and
on-shell supersymmetric cases differ from each other only by a $\B$-exact term. This is a consequence of the fact that
the BRST cohomology of both cases are the same. The missing terms that are found by climbing up with pure on-shell
supervariations \cite{ulker1} are then restored with the help of this $\B$-exact term.

\section{N=2 SYM theory, extended BRST transformations\\ and Slavnov-Taylor operator}

In this work we study  the formulation of N=2 SYM theory given in \cite{gsw,soh} by using the supersymmetry conventions
of \cite{wb}. We begin our analysis with the off-shell supersymmetric Yang-Mills theory in Wess-Zumino gauge. The action
of the theory
\bea
S_{N=2} &=& \frac{1}{g^2}Tr \int d^4 x (-\frac{1}{4} F_{\mu\nu} F^{\mu\nu}  -i\la^i D \!\!\!\! / \labar_i  + \f D_{\mu}
D^{\mu} \f^{\dag}\nonumber\\
&&\qquad\qquad\qquad  -\frac{i\kok}{2}(\la_i [\la^i , \ft ] +\labar^i [\labar_i ,\f ]) -\yarim [\f ,\ft]^2 +
\yarim\vD .\vD)
\eea
contains fields of N=2 vector multiplet $V=(A_\mu\, ,\,\f \, ,\, \ft \, ,\, \la_{i\a} \, ,\, \labar^{i}_{\adot} \, ,\,
\vD )$ , where  the gauge field $A_\mu  $ and the scalar fields $\f\, ,\,\ft $ are singlets, the Weyl spinors $\la_{i\a}
\,\, \labar^{i}_{\adot}$ are doublets and the auxiliary field $\vD $ is a triplet under the $SU(2)_R$ symmetry group
\cite{gsw,soh}. The  $SU(2)_R$ indices of the spinors are raised and lowered due to
\be
\la^i = \se^{ij} \la_j \quad ,\quad \la_i = \la^j \se_{ji} \quad ,\quad
\labar_i = \se_{ij} \labar^j \quad ,\quad \labar^i = \labar_j \se^{ji}
\ee
where the antisymmetric tensor $\se^{ij}$ is given as\footnote{Note that in our convention the $\se^{ij}$ is different
then the one, $\e_{\a\b}$,  used for spinor indices},
$$
\se_{12} = \se^{12} = -\se_{21} = - \se^{21}=1.
$$
The action (1) is invariant under N=2 supersymmetry transformations,
$$
\d = \t^{i\a} \qu _{i\a} + \tbar_{i \adot} \qubar ^{i\adot}
$$
that obey the following off-shell algebra
\bea
\{ \qu_i  ,\qubar^j \}  &=& -2i\d_i ^j \s^{\mu} D_{\mu} \nonumber\\
\{  \qu _{i\a} ,  \qu _{j\b} \} = -2i\kok \se_{ij} \e_{\a\b} \d_g (\ft) &,& \{\qubar ^{i\adot} , \qubar ^{j\bdot} \}
=-2i\kok \se^{ij} \e^{\adot \bdot} \d_g (\f).
\eea
where $\qu _{i\a}$ and $\qubar ^{i\adot}$ are chiral and antichiral part of the supersymmetry transformations,
\(\t^i , \tbar_i \) are corresponding anti-commuting supersymmetry parameters, and  $\d_g$ denotes field
dependent gauge transformations with respect to its argument.

The extension of BRST transformations to include global symmetries is well known \cite{w,mag,mpw,bra1}
 and is first given in Ref.\cite{mag} for N=2 supersymmetry. Following the standard procedure, on the members of N=2
vector multiplet such an extended BRST generator $s$ can be defined as
\be
s :=s_0 -i \x^i \qu_i -i \xbar_i \qubar^i
\ee
where $s_0$ is the ordinary BRST transformations and  $\x^{i\a}$ and $ \xbar_{i\adot} $  are the constant commuting
chiral and antichiral SUSY ghosts respectively. Note that since $s_0$ carries ghost number and it is anti-commuting, the
parameters of the global supersymmetry transformations are promoted to the status of constant ghosts and their Grassmann
parity are changed so that the  extended transformation $s$ is a homogeneous transformation.

The extended BRST transformation of the fields can now be written as,
\bea
sA_{\mu}&=& D_{\mu}c +\x_i \s_{\mu}\labar^i + \xbar^i \sbar_{\mu}\la_i \\
s\la_i &=& i\{c,\la_i \} - i\s^{\mu\nu}\x_i F_{\mu\nu} +\x_i [\f ,\ft ]  -\kok \s^{\mu}\xbar_i D_{\mu} \f + \vt_i ^j
\x_j.\vD \\
s\labar^i &=& i \{ c,\labar^i \} - i \sbar^{\mu\nu} \xbar^i F_{\mu\nu} -\xbar^i [\f ,\ft ]  -\kok\sbar^{\mu}\x^i D_{\mu}
\ft -\xbar^j \vt_i ^j . \vD   \\
s\f &=& i[c,\f ]-i\kok \x_i\la^i \\
s\ft &=& i[c,\ft ]-i\kok \xbar^i \labar_i \\
s\vD &=& i[c, \vD] + i\vt_i ^j(\x_j \Ds\labar^i -\xbar^i\bar{\Ds}\la_j  +\kok\x^i [\la_j ,\ft] -\kok\xbar_j
[\labar^i,\f]) \\
sc &=& \frac{i}{2}\{ c,c\}-2i\x_i\s^{\mu}\xbar^i A_{\mu} -\kok\x_i\x^i \ft -\kok\xbar^i\xbar_i \f
\eea
where $c$ is the usual Faddeev-Popov ghost field and  $\vt$'s are Pauli spin matrices. Note that with the help of extra
terms in $sc$, $s^2$ closes on translations,
\be
s^2 = -2i \x^i \s^\nu \xbar_i \del_\nu
\ee
in other words the complication that SUSY algebra is modified by field-dependent gauge transformations is
solved. Note also that in order to get a nilpotent $s$, the definition (4) could be extended to include translations
by introducing  suitable translation ghosts (see for instance \cite{mag}). However, since our aim is to work with the
integrated field polynomials the definition of $s$ given in (4)  will cause no problems.

Since the action (1) is invariant under gauge transformations and supersymmetry, it is obviously invariant under
extended BRST transformation $s$. The gauge fixing of the action (1) can be performed by adding an $s$-exact term
\cite{psbook}, that is compatible with supersymmetry, since the extended BRST operator $s$ contains supersymmetry. We
choose this term to be Landau type,
\be
S_{gf} = -tr \int d^4 x s(\cbar \del^\mu A_\mu)
\ee
where the fields $(\cbar, b)$ are the trivial pair that are introduced in the standard procedure of gauge fixing,
\be
s\cbar = b \quad ,\quad sb= -2i\x \s^\nu \xbar \del_\nu \cbar  .
\ee

In order to write Slavnov-Taylor (ST) identity from the gauge fixed action, the field content of the theory
is extended to include antifields (sources) that couple to the corresponding $s$-transformations of the fields,
\be
S_{quad}= tr\intx (A_{\mu}^* sA^{\mu} +\f^* s \f +\f^{\dag *} s\ft + \la^{*i}s\la_i + \labar_i ^* s\labar^i + \vD^* s\vD
+c^* sc)
\ee

For off-shell supersymmetric case, there is no need to extend the action further. The total action is now given by,
\be
I = S_{N=2} + S_{gf} + S_{ext}
\ee
and satisfies the following ST identity,
\bea
\S(I)&=& tr \int d^4 x ( \frac{\d I}{\d \fa}\frac{\d I}{\d \fas}+ s\cbar \frac{\del I}{\del \cbar} + sb
\frac{\del I}{\del b}) \\
   &= & 2i \x ^i \s^\nu \xbar_i  \, \Delta_\nu
\eea
where
$$\fa = \{A_{\mu},\la_{i\a} ,\labar^{i\adot},\f,\ft,\vD ,c\}\quad,\quad
\fas = \{A_{\mu}^*,\la^{* i\a },\labar_{i\adot}^*,\f^*,\f^{\dag*},\vD^*,c^*\} .$$ Here, $\Delta_\nu$ is a classical
breaking\footnote{ $(-1)^A$ denotes the Grassman parity of the field $\fa$ . },
\be
\Delta_{\nu} = tr\int d^4 x( (-1)^A \fas\del_\nu \fa )
\ee
due to the fact that, the translation symmetry is not included in $s$. It is well known that this classical breaking has
no effect on the renormalization of the theory, since it is linear in the fields \cite{psbook,sorlec}. Note that with
help of the extended BRST generator $s$, Ward identities for supersymmetry  are transformed into a unique ST identity
that includes all these symmetries and the identity (18) can be used to analyze the renormalization of N=2 SYM theory
\cite{psbook,mag}.

The so called linearized ST operator $\B$ \cite{psbook,sorlec}, that is the relevant object for cohomological analysis,
can be obtained from (17) as,
\be
\B_I =tr \int d^4 x ( \frac{\d I}{\d \fa}\frac{\d }{\d \fas}+ \frac{\d I}{\d \fas}\frac{\d }{\d \fa}+s\cbar \frac{\del
}{\del \cbar} + sb \frac{\del }{\del b}).
\ee
and it satisfies,
\be
\B_I \B_I = -2i\x \s^\nu \xbar \P_\nu
\ee
where $\P_\nu =  \int d^4 x ( \del_\nu \fa \frac{\del}{\del \fa} + \del_\nu \fas \frac{\del}{\del \fas})$ is a total
derivative when it is acted on the space of integrated field polynomials and therefore $\B$ can be considered as a
nilpotent operator on this space.

The on-shell supersymmetric N=2 SYM theory  is obtained , as usual, by eliminating the auxiliary field $\vD$ with its
equation of motion, that is $\vD=0$ for pure N=2 SYM theory. The on-shell action
 $\tilde S_{N=2} =S_{N=2}\vert_{\vD=0}$ is still invariant under on-shell supersymmetry transformations, but the
supersymmetry algebra (3) is satisfied only when the equations of motion of the spinor fields are used. As a
consequence, the extended BRST transformation that contains on-shell supersymmetry transformations
$$
\tilde{s} = s \vert _{\D=0}
$$
satisfies
\be
\tilde{s}^2 = -2i \x^i \s^\nu \xbar_i \del_\nu \qquad\mbox{(modulo eq. of motion  of  $\la \, , \labar$)} .
\ee

However, this complication that the algebra is modified by modulo terms involving equations of motion of spinor fields
can be rectified by adding a quadratic term in the anti-fields to the extended action (16) and thus a Slavnov-Taylor
operator that also  squares to a boundary term can be obtained \cite{w,mag,mpw,bra1}. For N=2 SYM it is found to
be:
\be
 S_{quad}= -\yarim g^2 tr\intx(\vt_i^j(\x_j\la^{*i}-\xbar^i \labar_j^*).\vt_k^l(\x_l\la^{*k}-\xbar^k \labar_l^*)).
\ee

The total on-shell classical action now reads,
\be
\tilde{I} = S_{N=2} \vert _{D=0}+S_{gf}+S_{ext}\vert _{\D=0}+S_{quad}
\ee
and following the same steps corresponding ST identity and the linearized ST operator can be obtained as,
\be
\S( \tilde{I} )=  2i \x \s^\nu \xbar \tilde{\Delta} _\nu
\ee
\be
\B_{\tilde{I}} =tr \int d^4 x ( \frac{\d \tilde{I}}{\d \tilde{\fa}}\frac{\d }{\d \tilde{\fas}}
+ \frac{\d \tilde{I}}{\d \tilde{\fas}}\frac{\d }{\d\tilde{\fa}}+s\cbar
\frac{\del }{\del \cbar} + sb \frac{\del }{\del b})
\ee
\be
\B_{\tilde{I}} \B_{\tilde{I}} = -2i\x \s^\nu \xbar \tilde{\P} _\nu
\ee
for $\tilde{\fa}= \fa\vert_{\D=0}\, ,\,    \tilde{\fas}=\fas\vert_{\D=0}\, ,\,
\tilde{\Delta} _\nu =\Delta_\nu \vert_{\D=0}  \, ,\,  \tilde{\P} _\nu =\P_\nu \vert_{\D=0}$.

Note that, ST identity (25) and ST operator (27) given for on-shell case are quite similar with ones of off-shell
case (18,21). This is due to the fact that  the combination
$$g^2 (\vt_i^j(\x_j\la^{*i}-\xbar^i \labar_j^*)$$
exactly behaves like the auxiliary field $\vD$,
\be
\B_{\tilde{I}} \la_i = \tilde{s}\la_i + g^2\vt_i^j \x_j.\vt_k^l(\x_l\la^{*k}-\xbar^k \labar_l^*) = \B_I \la_i
|_{\vD=\vt_k^l(\x_l\la^{*k}-\xbar^k \labar_l^*)}
\ee
\be
\B_{\tilde{I}} \labar^i = \tilde{s}\labar^i - g^2\xbar^j \vt_j^i .\vt_k^l(\x_l\la^{*k}-\xbar^k \labar_l^*) = \B_I
\labar^i |_{\vD=\vt_k^l(\x_l\la^{*k}-\xbar^k \labar_l^*)}
\ee
\be
\B_{\tilde{I}} g^2 (\vt_i^j(\x_j\la^{*i}-\xbar^i \labar_j^*) =  \B_I \vD |_{\vD=\vt_k^l(\x_l\la^{*k}-\xbar^k
\labar_l^*)}
\ee
It is worth underlining that, a similar relation between the auxiliary field and the certain combination of the
supersymmetry ghosts and antifields of spinor fields also exists for N=1 SYM theory \cite{bra3,ulker2}. It seems natural
that, the quadratic terms in antifields of spinor fields that has to be added to the on-shell action  in order to obtain
a nilpotent operator, should be related to the auxiliary fields of  SYM theories. Indeed, in BV formalism such combinations of anti-fields arise naturally when one eliminates the auxiliary fields using their 'generalized equations of motion' which are derived from the master action rather than from the classical action \cite{bra3,hen}. It can be interesting to find out if this relation can be used to obtain an off-shell formulation of the supersymmetric theories where the auxiliary field content is not known, such as N=4 SYM theory.

\section{Descent equations for N=2 SYM theory}

As it is discussed in the previous section, when the linearized ST operator\footnote{For the following discussion, $\B$
will stand for both $\B_I$ and $\B_{\tilde{I}}$.} $\B$ is defined on the space of integrated polynomials of fields and
antifields, it is nilpotent and it constitutes a cohomology problem on this functional space,
\be
\B\intx \X = 0
\ee
where the physically intresting solutions are the ones that can not be written as a $\B$-exact term,
$$ \intx\X\not=B\intx\X'  .$$

One way of characterizing the cohomology classes of the operator $\B$ is to study the set of descent equations stemming
from Wess-Zumino consistency condition. This framework is also the relevant one for our purposes since our aim is to
relate the action of N=2 SYM with lower dimensional field polynomials.

To derive the set of descent equations we will generalize the strategy given in Ref.\cite{ulker2} to N=2 supersymmetric
case. The first descent equation is obtained, as usual,  by taking the local version of
eq.(31):
\be
\B\X^{(0)} = \xbar^i_{\adot} \sbar^{\mu\adot\a}\del_\mu \X^{(1)}_{i\a}
\ee
Here, the $\xbar^i_{\adot} \sbar^{\mu\adot\a}$ factor  is assumed to appear due to supersymmetry algebra. To derive the
second of the descent equations we apply $\B$ to Eq.(32)
$$ \B^2 \X^{(0)} = -2i \xbar^i \sbar^\mu \x_i \del_\mu \X^{(0)} = \xbar^i_{\adot} \sbar^{\mu\adot\a}\del_\mu \B
\X^{(1)}_{i\a} $$
that implies
\be
\xbar^i_{\adot} \sbar^{\mu\adot\a}\del_\mu  (2i  \x_{i\a} \X^{(0)} + \B \X^{(1)}_{i\a} ) = 0.
\ee
From the condition (33) the second descent equation can be found as,
\be
\B\X^{(1)}_{i\a} = -2i \x_{i\a}\X^{(0)} + (\xbar^j \sbar^\mu )^{\b}\del_\mu \X^{(2)}_{i\a,j\b}
\ee
where $\X^{(2)}_{i\a,j\b}$ is a local polynomial antisymmetric in the pair of indices $(i\a)$ and $(j\b)$, i.e.
$\X^{(2)}_{i\a,j\b}= -\X^{(2)}_{j\b,i\a}$. Note that the antisymmetry of $\X^{(2)}$ follows from the fact that the term
$$
\xbar^i _{\adot} \sbar^{\adot\a\mu}\del_\mu\, \xbar^i_{\bdot} \sbar^{\bdot\b\nu}  \del_\nu \X^{(2)} _{i\a,j\b}
$$
that can be added to condition (33), vanishes due to the commuting nature of the global supersymmetry ghosts only when
 $\X^{(2)}$ is taken to be antisymmetric in the pair of its indices. The rest of the descent equations can be found
easily by iterating this procedure and the following set of descent equations for N=2 supersymmetry can be
written:
\bea
\B\X^{(0)} &=& \xbar^i_{\adot} \sbar^{\mu\adot\a}\del_\mu \X^{(1)}_{i\a}\\
\B\X^{(1)}_{i\a} &=& -2i \x_{i\a}\X^{(0)} + (\xbar^j \sbar^\mu )^{\b}\del_\mu \X^{(2)}_{i\a,j\b}\\
\B\X^{(2)}_{i\a,j\b} &=& -2i \x_{j\b}\X^{(1)}_{i\a}+2i \x_{i\a}\X^{(1)}_{j\b} + (\xbar^k \sbar^\mu )^{\g}\del_\mu
\X^{(3)}_{i\a,j\b,k\g}\\
\B\X^{(3)}_{i\a,j\b,k\g} &=& 2i \x_{j\b}\X^{(2)}_{i\a,k\g} -2i \x_{i\a}\X^{(2)}_{j\b,k\g}
-2i\x_{k\g}\X^{(2)}_{i\a,j\b} +(\xbar^l \sbar^\mu )^{\la}\del_\mu  \X^{(4)}_{i\a,j\b,k\g,l\la}\\
\B\X^{(4)}_{i\a,j\b,k\g,l\la}&=& -2i\x_{l\la}\X^{(3)}_{i\a,j\b,k\g}-2i\x_{j\b}\X^{(3)}_{i\a,k\g,l\la} +
2i\x_{i\a}\X^{(3)}_{j\b,k\g,l\la} + 2i\x_{k\g}\X^{(3)}_{i\a,j\b,l\la}
\eea
Here, the local polynomials $\X^{(3)}$, $\X^{(4)}$ are also totally antisymmetric in the pair of spinor and SU(2)-R
indices (like when the pairs $(i\a),(j\b),...$ are exchanged). The descent equations terminate at the fourth level, due
to the fact that the pair of spinor and SU(2)-R indices can take only four distinct values, i.e. a totally antisymmetric
$\X^{(5)}$ is zero identically. Therefore, when these equations are compared with that of N=1 SYM \cite{ulker2}, it is
clear that the number of descent equations are related directly with the internal R-symmetry content of the theory.

The structure of descent equations (35-39) for N=2 SYM theory are quite different and complicated from the ones for
standard gauge theories (see for instance Ref.s\cite{psbook,bbhrev}). Due to the assumption that the antichiral
supersymmetry ghosts, $\xbar^i$, appear explicitly infront of the derivatives on the R.H.S. of the descent equations,
all the solutions $\X^{(i)}$ carry the same ghost number and   the possible solutions are highly constrained. This is
due to the supersymmetric structure of the theory. Moreover, the last equation is not homogeneous. Nevertheless,  the
RHS of the last equation is homogeneous in $\xbar^i$  and therefore, introduction of a filtration of the linearized ST
operator with respect to chiral ghosts $\x^i$,
\be
\mathcal{N} = \x^{i\a} \frac{\d}{\d \x^{i\a}}\quad ;\quad \B = \sum{\B_n} \quad ,\quad [\mathcal{N}, \B_n] = n \B_n
\ee
which leads to the algebra
\bea
\B_0 ^2 &=&0\\
\{\B_0,\B_1\}&=&-2i\x \s^\mu \xbar \del_{\mu}\\
\{\B_0,\B_2\}+\B_1 ^2 &=&0\\
\B_2 ^2 = \{\B_1,\B_2\}&=&0.
\eea
is useful to find a solution.

Due to the filtration (40), the lowest descent equation (39) can be divided into two,
\be
\B_0 \,  \X^{(4)}_{i\a,j\b,k\g,l\la} = 0
\ee
\be
\x^i \q_{i\a}\, \X^{(4)}_{i\a,j\b,k\g,l\la} = -2i\x_{l\la}\X^{(3)}_{i\a,j\b,k\g}-2i\x_{j\b}\X^{(3)}_{i\a,k\g,l\la} +
2i\x_{i\a}\X^{(3)}_{j\b,k\g,l\la} + 2i\x_{k\g}\X^{(3)}_{i\a,j\b,l\la}
\ee
where we have defined the operator $\q_{i\a}$ as
\be
\x^i \q_{i\a} = \B_1 + \B_2.
\ee

Note that since the zeroth order operator $\B_0$ in the filtration of $\B$ is strictly nilpotent, it also constitutes  a
cohomology problem and  the cohomology of the full operator $\B$ is isomorphic to a subspace of the cohomology of the
operator $\B_0$ \cite{psbook,sorlec}.

On the other hand, the equation (45,46) indicates that the operator $\q_\a$ can be used as a kind of climbing up
operator. Indeed, whenever the descent equations can be devided into two like Eq.s(45,46),  after some algebra it is
found that the solutions of descent equations (35-39) are algebraically related to each other as,

\bea
\q^{i\a}\q^{j\b}\q^{k\g}\q^{l\la} \, \X^{(4)}_{i\a,j\b,k\g,l\la} &=& (2i)^4 4!\, \X^{(0)} \nonumber\\
\q^{j\b}\q^{k\g}\q^{l\la}\, \X^{(4)}_{i\a,j\b,k\g,l\la} &=& (2i)^3 3!\, \X^{(1)}_{i\a} \nonumber\\
\q^{k\g}\q^{l\la}\, \X^{(4)}_{i\a,j\b,k\g,l\la} &=& (2i)^2 2!\, \X^{(2)}_{i\a,j\b} \nonumber\\
\q^{l\la}\, \X^{(4)}_{i\a,j\b,k\g,l\la} &=& \,(2i)\, 1!\, \X^{(3)}_{i\a,j\b,k\g}.
\eea

Therefore, once a explicit solution of the lowest descent equation is found, that belongs to the cohomology of
the filtered operator $\B_0$, the higher solutions in the descent can be obtained by applying the climbing up operator
$\q_{i\a}$ both for off- and on-shell supersymmetric cases. Note that  the framework given above in order to solve the
cohomology problem of the linearized ST operator $\B$ is a direct generalization of the method presented in
Ref.\cite{ulker2} for N=1 SYM theory and as expected, it is smiliar with the ones given for the twisted N=2 SYM
\cite{sorella,sorlec,sorella2}.

\section{Construction of the N=2 SYM action}

In algebraic renormalization framework the solutions of the cohomology problem (31) determined by the linearized ST
operator $\B$ for ghost numbers 0 and 1 gives the invariant counterterms that can be added to any order in the
perturbation theory and the possible anomalies respectively. As a consequence the classical action also belongs to the
cohomology of $\B$ in the ghost sector zero and  to study the descent equations for analyzing the cohomology of $\B$ of
a supersymmetric theory gives a natural framework to relate the corresponding action to the lower dimensional field
polynomials.

Therefore, we are intrested in a gauge invariant solution $\X^{(0)}$ of Eq.(31) , which has the same quantum numbers
with the classical Lagrangean of the N=2 SYM theory, i.e. a solution that  has dimension four with vanishing ghost
number and $SU(2)$-R charge and Grassmann even. The solutions of the lower descent equations are also constrained due to
this requirement and as a consequence the gauge invariant solution $\X^{(4)}$ of the lowest descent equation (39) has
dimension 2, and R-charge -4 (see table 1).
\begin{table}[hbt]
\caption[t1]{ Dimensions   $d$, Grassmann parity $GP$, ghost number $Gh$  and R-weights.}
\centering
\begin{tabular}{|c||c|c|c|c|c|c|c|c|c|c|c|c|c|c|}
\hline
&$A_\mu$&$\la_i$&$\f$&$\vD$&$c$&$\cbar$&$b$&$\x_i$&$A^*_\mu$&$\la_i^*$&$\f^*$&$\vD^*$&$c^*$&$\B$ \\
\hline\hline
$d$&1&3/2&1&2&0&2&2&-1/2&3&5/2&3&2&4&0 \\
\hline
$GP$&0&1&0&0&1&1&0&0&1&0&1&1&0&1 \\
\hline
$Gh$&0&0&0&0&1&-1&0&1&-1&-1&-1&-1&-2&1\\
\hline
$R$&0&-1&-2&0&0&0&0&-1&0&1&2&0&0&0 \\
\hline
\end{tabular}
\end{table}

On the other hand, as discussed in the previous section, the solution $\X^{(4)}$ has to belong to the cohomology of the
filtered operator $\B_0$,
$$\B_0 \X^{(4)} = 0 $$
Note that this condition also implies that the solution $\X^{(4})$ is gauge invariant. The only such gauge invariant
field polynomial with correct quantum numbers is $Tr\f^2$ for both off- and on-shell supersymmetric cases
and the solution $\X^{(4)}$ with desired index structure to the descent equation (39) can be written as,
\be
\X^{(4)}_{i\a,j\b,k\g,l\la}=k\, E_{i\a,j\b,k\g,l\la}\, Tr ( \f^2)
\ee
where $k$ is a constant and $E_{i\a,j\b,k\g,l\la}$ is a totally anti-symmetric tensor in the pair of indices
$(i\a),(j\b),(k\g),(l\la)$,
\be
E_{i\a,j\b,k\g,l\la} = \frac{2}{3} (\, \e_{\a\b}\e_{\g\la}(\se_{ik}\se_{jl}+\se_{il}\se_{jk}) -
\e_{\a\g}\e_{\b\la}(\se_{ij}\se_{kl}+\se_{il}\se_{kj})
 + \e_{\a\la}\e_{\b\g}(\se_{ij}\se_{lk}+\se_{ik}\se_{lj} ).
\ee

It is worthwhile to remark that the field monomial $Tr\f^2$ is a nontrivial element of the cohomology of operator $\B_0$
 only when  $\B_0$  is defined on the space of field polynomials that are analytic in constant supersymmetry
 ghosts. In other words, if the functional space is defined to be the polynomials of the fields that are not necessarily
analytic in the constant ghosts $\bar{\x}^i$, $Tr\f^2$ can be written  as an exact-$\B_0$ term,
$$
Tr\f^2 = \B_0 \, Tr(-\frac{\kok}{2\xbar^i\xbar_i}c\f +\frac{i\kok}{12(\xbar^i\xbar_i)^2}\{ c,c\} c)
$$
and therefore it belongs to the trivial cohomology of $\B_0$. This fact has also been pointed out in Ref.s
\cite{sorella,sorlec}, for twisted formulation of N=2 SYM that the twist of the N=2 theory can be interpreted as a
topological theory only if the analyticity is lost in (scalar) SUSY ghosts. Moreover, in a recent paper a smiliar
non-analyticity argument in constant supersymmetry ghosts is used to show that the topological Yang-Mills (TYM)
\cite{witten} theory can be obtained by using field redefinitions i.e. as a change of variables (without twisting)
\cite{ulker3}. Therefore, physical and topological interpretations of N=2 SYM are intertwined together due to the
requirement of analyticity of supersymmetry ghosts\footnote{As noted in Ref.\cite{sorella,sorlec},  when perturbative
calculations are considered one should obviously require analyticity of the parameters of a theory, the analyticity
requirement for supersymmetry ghosts is mandatory for perturbative regime. } \cite{sorella,sorlec,ulker3}.

It is  straight forward to find the solution $\X^{(0)}$ from the lowest solution $\X^{(4)}$ (49) by using the lift given
in (40) for both off-shell supersymmetric case,
\bea
\X^{(0)}_{off} &=& k Tr  (-\frac{1}{4} F_{\mu\nu} F^{\mu\nu} -\frac{i}{8} \epsilon ^{\mu\nu\la\k}F_{\mu\nu}F_{\la\k}
 -i\la^i D \!\!\!\! / \labar_i  + \f D_{\mu}
D^{\mu} \f^{\dag}  \nonumber\\
&&\qquad\qquad\qquad -\frac{i\kok}{2}(\la_i [\la^i , \ft ] +\labar^i [\labar_i ,\f ]) -\yarim [\f ,\ft]^2 +
\yarim\vD .\vD)
\eea
and for on-shell supersymmetric case
\be
\X^{(0)}_{on} = k Tr  (-\frac{1}{4} F_{\mu\nu} F^{\mu\nu} -\frac{i}{8} \epsilon ^{\mu\nu\la\k}F_{\mu\nu}F_{\la\k}
 -\frac{i}{2}\la^i D \!\!\!\! / \labar_i
+ \yarim \f D_{\mu}D^{\mu} \f^{\dag}  -\frac{i\kok}{4}\labar^i [\labar_i ,\f ]) + \mathcal{F} (\fa, \fas)
\ee
where $\mathcal{F} (\fa, \fas)$ is a complicated polynomial of fields and anti-fields.

Note that in the above solutions the term $ Tr \epsilon ^{\mu\nu\la\k}F_{\mu\nu}F_{\la\k}$ is a total derivative under
the integral sign and it is seen easily that  off-shell (extended) N=2 SYM action is  directly related to the lower
dimensional field monomial $Tr \f^2$ as a consequence of the lift (48) for $k=\frac{1}{g^2}$,
\be
-\yarim g \frac{d}{dg} I =  S_{N=2} = \frac{1}{g^2} \q^4 \, Tr \int d^4 x \f^2
\ee
where we have used $\q^4 = \frac{1}{2^4 4!}E_{i\a,j\b,k\g,l\la} \q^{i\a}\q^{j\b}\q^{k\g}\q^{l\la}$ for notational
simplicity.

The on-shell (extended) action can be obtained, by noting that the anti-field independent part of the solution differs
from the original action by equations of motion terms due to on-shell supersymmetry. Since, $\B$-transformation of
antifields includes the equation of motion of the corresponding field\footnote{This part of the operator
$\B$ is often called in the literature Koszul-Tate differential. See Ref. \cite{bbhrev} for its importance for BRST
cohomological calculations.},
$$ \B_{\tilde{I}} \fas = \frac{\d \tilde{I} }{\d \fa } = \frac{\d S_{N=2} }{\d \fa } + ...$$
addition of a $\B_{\tilde{I}}$-exact term restores these missing terms and the on-shell (extended) action can be
constructed as
\be
 -\yarim g \frac{d}{dg} \tilde{I} = S_{N=2}\vert _{\D =0} - S_{quad} =  \frac{1}{g^2} \q^4 \,Tr\intx \f^2 +
\frac{1}{2}\B_{\tilde{I}}\, tr\int d^4 x (\f^* \f - \la^{*i} \la_i ) .
\ee

The above relations show that both off and on-shell supersymmetric SYM actions can be constructed from 2 dimensional
field monomial $Tr \f^2$. The difference between two cases is only a $\B$-exact term, which can be interprated as a
result of  the theorem that  local BRST cohomologies of two formulations of the same theory differing in auxiliary field
content are the same \cite{bbh}. Moreover, since $Tr\f^2$ is the lowest component of the N=2 multiplet where the action
belongs, the relations (53) and (54) imply that the above method of working the BRST cohomology through the descent
equations, gives the relation between the action and the lowest component of the multiplet in an elegant way for
both off and on-shell supersymmetric cases. A similar structure is also obtained for N=1 SYM theory \cite{ulker2} and
 it should be straightforward to generalize this method to other supersymmetric theories.

\vskip2em
{\bf Acknowledgements: }
The author is grateful to R. Flume for introducing the subject. He also gratefully acknowledges the enlightening
discussions with \"{O}. F. Day\i{} and M. Horta\c{c}su.\\This work is partially supported by TUBITAK, Scientific and
Technical Research Council of Turkey, under BAYG-BDP.

\vskip3em

%
%
%
%

\end{document}